\documentclass[twocolumn]{aastex63}
\usepackage{graphicx}
\usepackage{booktabs}
\usepackage[figuresright]{rotating}

\begin{document}
\shorttitle{Deciphering Star Cluster Evolution}

\shortauthors{Hu, Q.S. et al.}

\title{Deciphering Star Cluster Evolution by Shape Morphology}

\author{Qingshun Hu}
\affil{Xinjiang Astronomical Observatory, Chinese Academy of Sciences, Urumqi, Xinjiang 830011, People's Republic of China}
\affil{University of Chinese Academy of Sciences, Beijing 100049, People's Republic of China}

\author{Yu Zhang}
\email{zhy@xao.ac.cn}
\affil{Xinjiang Astronomical Observatory, Chinese Academy of Sciences, Urumqi, Xinjiang 830011, People's Republic of China}
\affil{University of Chinese Academy of Sciences, Beijing 100049, People's Republic of China}

\author{Ali Esamdin}
\email{aliyi@xao.ac.cn}
\affil{Xinjiang Astronomical Observatory, Chinese Academy of Sciences, Urumqi, Xinjiang 830011, People's Republic of China}
\affil{University of Chinese Academy of Sciences, Beijing 100049, People's Republic of China}

\author{Jinzhong Liu}
\affil{Xinjiang Astronomical Observatory, Chinese Academy of Sciences, Urumqi, Xinjiang 830011, People's Republic of China}
\affil{University of Chinese Academy of Sciences, Beijing 100049, People's Republic of China}

\author{Xiangyun Zeng}
\affil{Xinjiang Astronomical Observatory, Chinese Academy of Sciences, Urumqi, Xinjiang 830011, People's Republic of China}
\affil{University of Chinese Academy of Sciences, Beijing 100049, People's Republic of China}

\begin{abstract}
We analyze the morphological evolution of open clusters and provide shape parameters for 265 open clusters. The results show that the overall shape of sample clusters becomes more elliptical as they grow older, while their core remains circular or slightly trend to circularize. There is a negative correlation of the ellipticities with the number of members of the sample clusters. A significant negative correlation between the overall ellipticities and masses is also detected for the sample clusters with log(age/year)~$\geq$~8, suggesting that the overall shapes of the clusters are possibly influenced by the number of members and masses, in addition to the external forces and the surrounding environment. For most young sample clusters, the radial stratification degree of the short axis direction is greater than that of the long, implying that the radial stratification degree in the two directions within the young sample cluster may be unevenly affected by an internal evolutionary process. Older sample clusters exhibit lower stratification in the tangential direction, which possibly means those clusters may continue to survive for a long time at a low level of stratification. Our analysis shows that the overall shape of the sample clusters may be more susceptible to the influence of Galactic tides toward the Galactic center than the shear forces embedded in Galactic differential rotation. By analyzing the distribution of the ages and number of members of star clusters, we suggest that NGC~6791 may originate from superclusters.
\end{abstract}

\keywords{Galaxy: stellar content --- methods: statistical --- open clusters and associations: general}

\section{Introduction}

Stars form in groups out of molecular clouds \citep[e.g.][]{lada03,alve12,mege16}. The youngest groups are distributed collectively as imprints of the usual filamentary cloud structures. Once surplus gas and dust is being dispersed \citep[e.g.][]{tutu78, hill80, elme83, math83, adam00, krou01, krou02, good06, moec10, pelu12, brin17}, two-body gravitational relaxation \citep[e.g.][]{berg01, ryon17, mamo19} among stellar members leads to sphericalization, i.e., an overall circular shape, with lower-mass members occupy larger volumes (``mass segregation''). The least-massive members are most vulnerable to dynamically ejection from the system (``stellar evaporation''), further loosening the gravitational binding \citep[e.g.][]{giel08, tang19}. External perturbation such as the tidal forces, differential rotation, or disk crossing, exacerbates the disintegration process, and eventually, the members would supply as Galactic field stars. The remnant systems, globular or open clusters we see today, with relatively well-determined distances, ages, and metallicities, serve to diagnose how stars of different masses evolve, and as laboratories to test stellar dynamics theories.

Open clusters, in particular, with dozens to thousands of member stars each \citep{berg01}, with ages from millions to billions of years, and located in the Galactic disk, provide a sample to probe the Galactic structure and evolution \citep{bala20}. Most star groups are likely dispersed upon emergence from molecular clouds \citep{lada03}, and the majority of the survival star clusters are expected to be dissolved in some $10^8$ years \citep[e.g.][]{wiel71, pand86, berg01, bona06, yang13}. Conceivably, longevity is possible only among systems relatively free of disturbances.

The dynamical evolution of an open cluster, arising from internal or external influences, manifests itself in its shape changes. As early as more than a century ago, the flattening of a moving cluster was reasoned \citep{jean16}, and such a phenomenon was indeed detected in the Hyades \citep{oort79}. \citet{pand90} identified the existence of ''corona'' features around open clusters. Such halo structures were characterized by \citet{nila02} based on the photometric data of the Digital Sky Survey. \citet{berg01} found open cluster to be generally elongated in a direction parallel to the plane, attributed to tidal forces. \citet{chen04}, with a sample of 31 open clusters from the 2MASS (Two Micron All Sky Survey) dataset, found clusters to be elongated even among the youngest systems, reflecting morphological relics in the parental molecular clouds; namely, young open clusters inherit the shape of the dense molecular cloud from which they formed \citep{getm18}. With the all-sky data of the compiled catalog of 2.5 million stars ASCC-2.5, \citet{khar09} derived the shape parameters, that is, the apparent ellipticity and orientation of the ellipse, of 650 open clusters and concluded that the ellipticities of open clusters generally range from 0.2 to 0.3, which are lower than those of modeled clusters. \citet{zhai17} computed the morphology parameters of 154 open clusters using the WEBDA database and suggested all clusters in their sample are elongated in shape. \citet{bhat17}, by using probabilistic star counting of Pan-STARRS point sources, detected the core-tail structure in the aged open cluster Berkeley~17, with the tails preferentially consisting of low-mass members.  With {\it Gaia} Data Release~2 (DR2) data, the substructure and halo of the double cluster $h$ and $\chi$ Persei was diagnosed by \citep{zhon19}, Similarly, tidal tails are found in the nearby clusters Coma Berenices \citep{tang19} and Blanco~1 \citep{zhan20}. The expected avalanche of high-precision kinematic and photometric data, particularly from {\it Gaia}, affords reliable membership determination in, and hence the studies of the dynamical evolution of, star clusters.

In this paper, we analyze a sample of 265 open clusters with which membership is available in the literature to diagnose the possible dominant processes in the history of the dynamical evolution of the open clusters in general. We present in Sec.~2 the data used in our study and the methodology to derive the shape parameters of the clusters. The results and the discussions are presented in Sec.~3. A summary of this work is given in Sec.~4.

\section{Data and Method}

The data used in this work is from the {\it Gaia} DR2 \citep[see, e.g.][]{gaia16, gaia18b}, a dataset of some 1.3 billion stars \citep[e.g.][]{gaia18b, aren18, cant18} with precise astrometry at the sub-milliarcsecond level \citep{lind18} and homogeneous photometry at the mmag level \citep{evan18}. The uncertainties of proper motion are about 0.05~mas~yr$^{-1}$ for bright sources (G $< 14$~mag) and 1.2~mas~yr$^{-1}$ for faint ones (G$\sim20$~mag). The uncertainties of parallax are about 0.03~mas and 0.7~mas for bright and faint sources, respectively \citep{lind18}. Reliable membership identification in the open clusters, essential in our work, is made possible by the high-quality {\it Gaia} data.

We started with the member catalog of open clusters \citep{cant18}, which provides the combined spatio-kinematic-photometric membership of 1229 open clusters based on the UPMASK method \citep{kron14} applied on the {\it Gaia} DR2. And, we adopted the age parameters of 269 open clusters from \citet{boss19}. After matching these two samples, we obtained 266 clusters in common. Except for Blanco~1, which is not included in our analysis because of its high Galactic latitude and too close a distance, the rest 265 clusters form the sample for our shape analysis, with logarithmic ages (years) ranging from 7.006 to 9.927, i.e., from relatively young to aged systems, fulfilling our requirement to diagnose the evolutionary of the open clusters.

We analyzed the shape of each cluster by using {\it Nonparametric Bivariate Density Estimation}, and {\it Least Square Ellipse Fitting}, both available in Python packages (Astropy, Matplotlib, Numpy, Scipy, and Sympy). Estimating the stellar number density profile according to the actual density distribution without model assumptions, the {\it Nonparametric Bivariate Density Estimation} is simple and nonparametric based on kernel density estimation. Its principle, in detail, is to estimate the probability density function of a random variable in a nonparametric way, which works for univariate and multivariate data. It has the advantage of delineating the actual density of the member stars in a cluster as far as possible. In this work, the kernel of the density estimation we adopted is Gaussian kernel. The shape parameters are then derived by the {\it Least Square Ellipse Fitting}, including the ellipticity $e$, and the position angle $q$ between the long principal axis of the ellipse and the Galactic plane. The value of ellipticity $e$, ranging between 0 and 1, represents the shape elongation, whereas $q$ refers to the orientation of a cluster in the disk, ranging from $-90\degr$ to $90\degr$.

\begin{figure*}[ht]
\centering
\includegraphics[width=110mm]{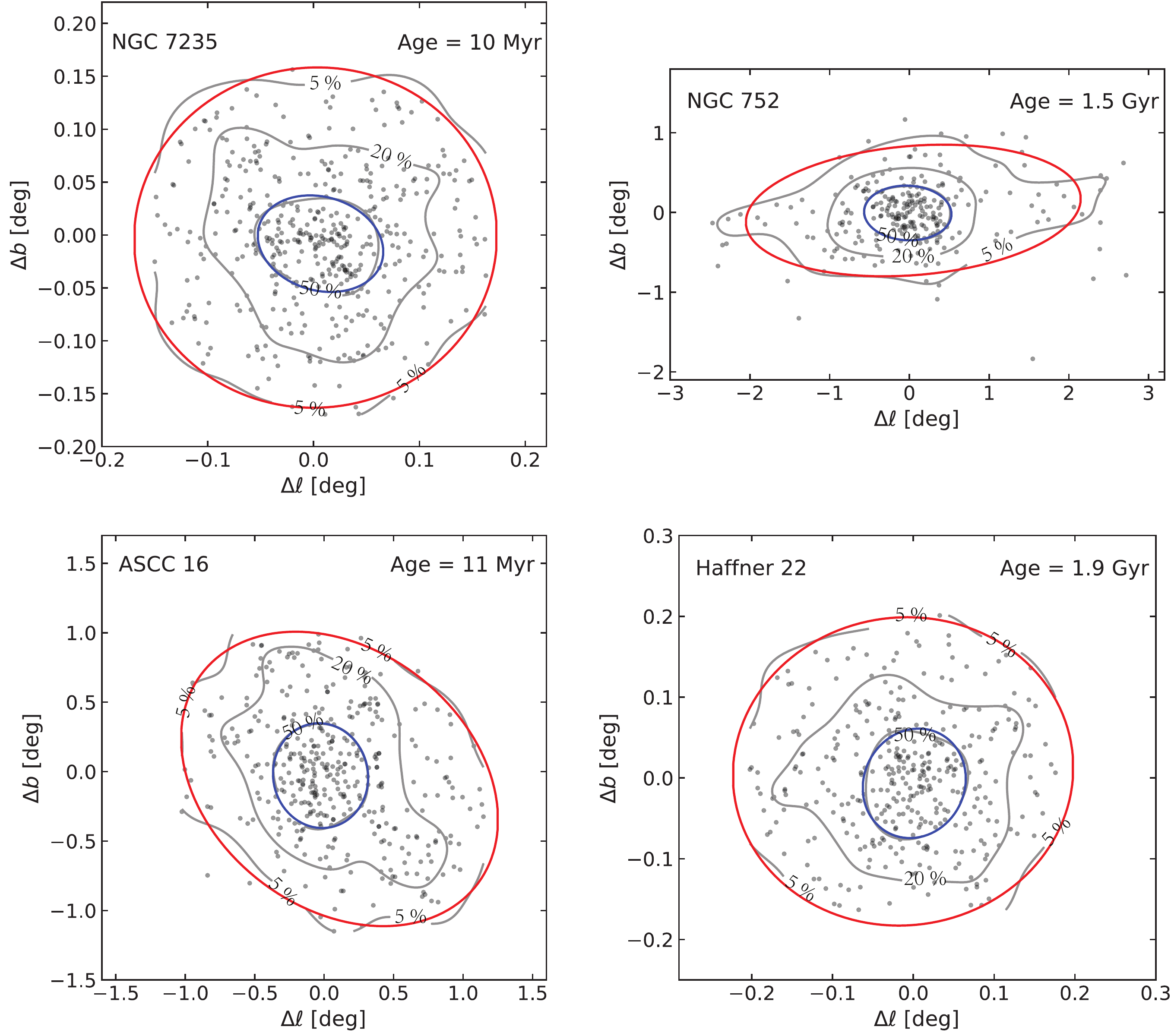}
\caption{The general scheme of fitting cluster density profiles for four open clusters (NGC~7235, NGC~752, ASCC~16, and Haffner~22). For each cluster in each panel, the black dots represent the member stars of a cluster with the gray lines as the stellar number density profiles at 50\% of and 5\% of the peak. the fitted ellipses of 50\% and 5\% of the peak stellar number density of the clusters are marked in blue and red, respectively. The percentage labels on the gray lines represent the relative percentage of the actual density value relative to the peak. The density contours at 5\% in the panel are incomplete because the cluster member stars show a non-dispersive distribution at the periphery of them.}
\label{fig:Density profile}
\end{figure*}

The shape parameters are derived in two steps. First, the members of each cluster with center $(l_0 , b_0)$ given in \citet{cant18} are plotted in a two-dimensional spherical Galactic coordinate system $(l , b)$. The density profile is then overplotted with the {\it Nonparametric Bivariate Density Estimation}. Figure~\ref{fig:Density profile} illustrates the results for four clusters, NGC~7235, NGC~752, ASCC~16, and Haffner~22. In estimating the densities, this work uses the maximum and minimum values of cluster member coordinate data as grid boundaries. The critical factor in the density profile is the grid size that dictates the degree of smoothness of the density profile.  Empirically we found the number of grid cells of 40000 as a satisfactory compromise between resolution and possible density fluctuations. Second, with the density profile, we assign two ellipses, via the {\it Least Square Ellipse Fitting}, to represent the shapes of the core and the overall of a cluster. For this, we set arbitrarily at 50\% and 5\% of the peak stellar number density of each open cluster. These ellipses are parameterized with $e_{core}$, $q_{core}$ for the inner shape, and $e_{all}$, $q_{all}$ for the overall shape of a cluster. Meanwhile, the additional parameters such as $a_{core}$ (half-length axes of inner ellipises), $a_{all}$ (half-length axes of overall ellipises), $b_{core}$ (half-short axes of inner ellipises), and $b_{all}$ (half-short axes of overall ellipises) are provided. We estimate the errors of ellipticities by using the standard deviation of the ellipse fitted. The median of relative errors of the inner ellipticities ($e_{core}$, corresponding to the inner shape of the clusters) for the sample clusters is about 4\%, and that of the overall ($e_{all}$, corresponding to the overall shape of the clusters) about 11\%. We note that one cluster, Platais~10, did not display a reliable solution with our analysis pipeline for the overall ellipticity due to the scattered distribution of its members; the overall shape, therefore, was estimated by visual inspection for it. The shape parameters and additional parameters of our sample are listed in Table~\ref{table:data1}.

\section{Results and discussion}

\begin{figure*}[ht]
\centering
\includegraphics[width=120mm]{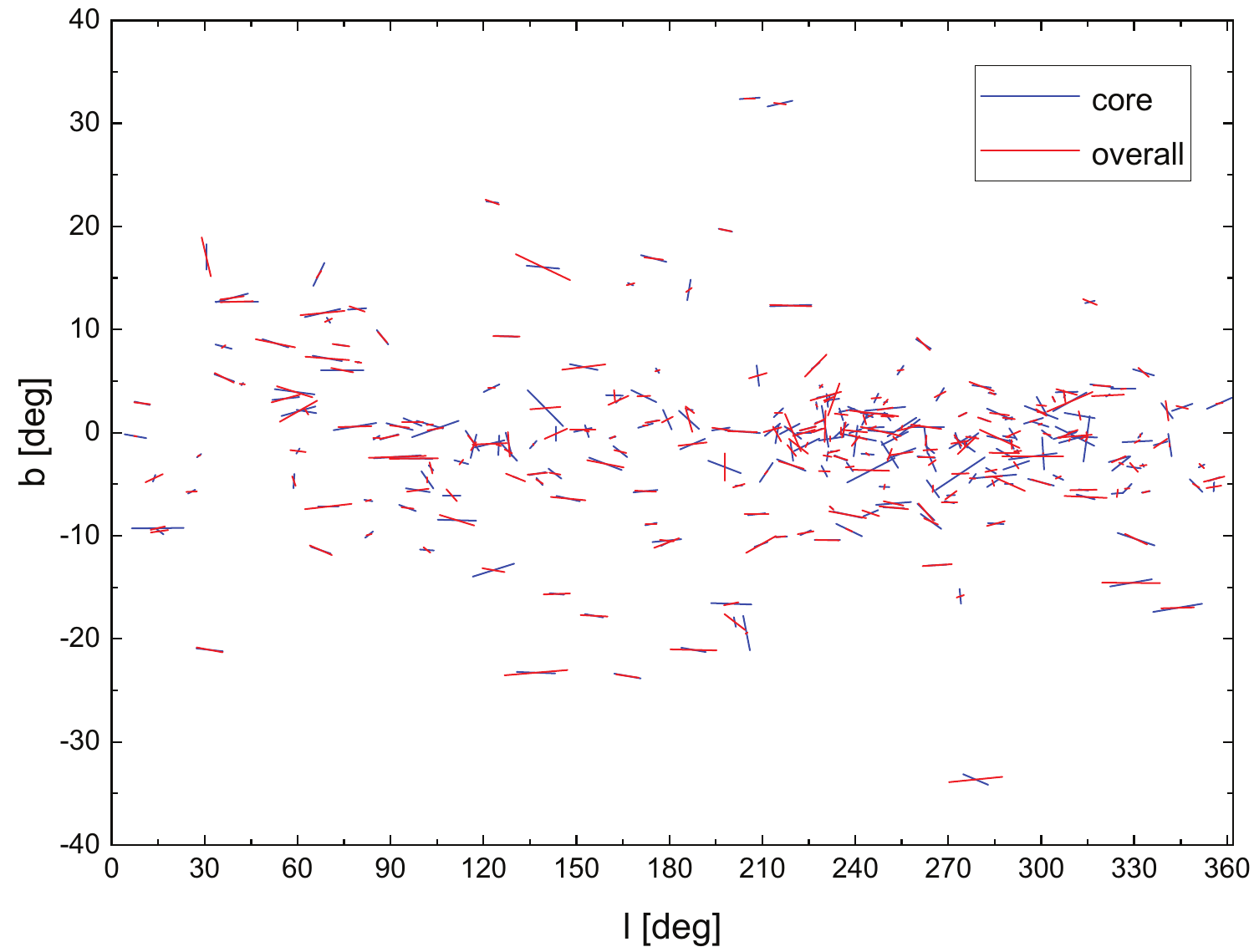}
\caption{The distributions of the sample clusters in Galactic coordinate system. The lengths of the blue and red bars are proportional to the ellipticities of the cores and overalls of the clusters, respectively. The inclinations of the blue and red bars indicate orientations of the cores and overalls of the clusters, respectively. The midpoints of the bars are set to the cluster's positions.}
\label{fig:Whole_sample}
\end{figure*}

\begin{figure}[ht]
\centering
\includegraphics[width=76mm]{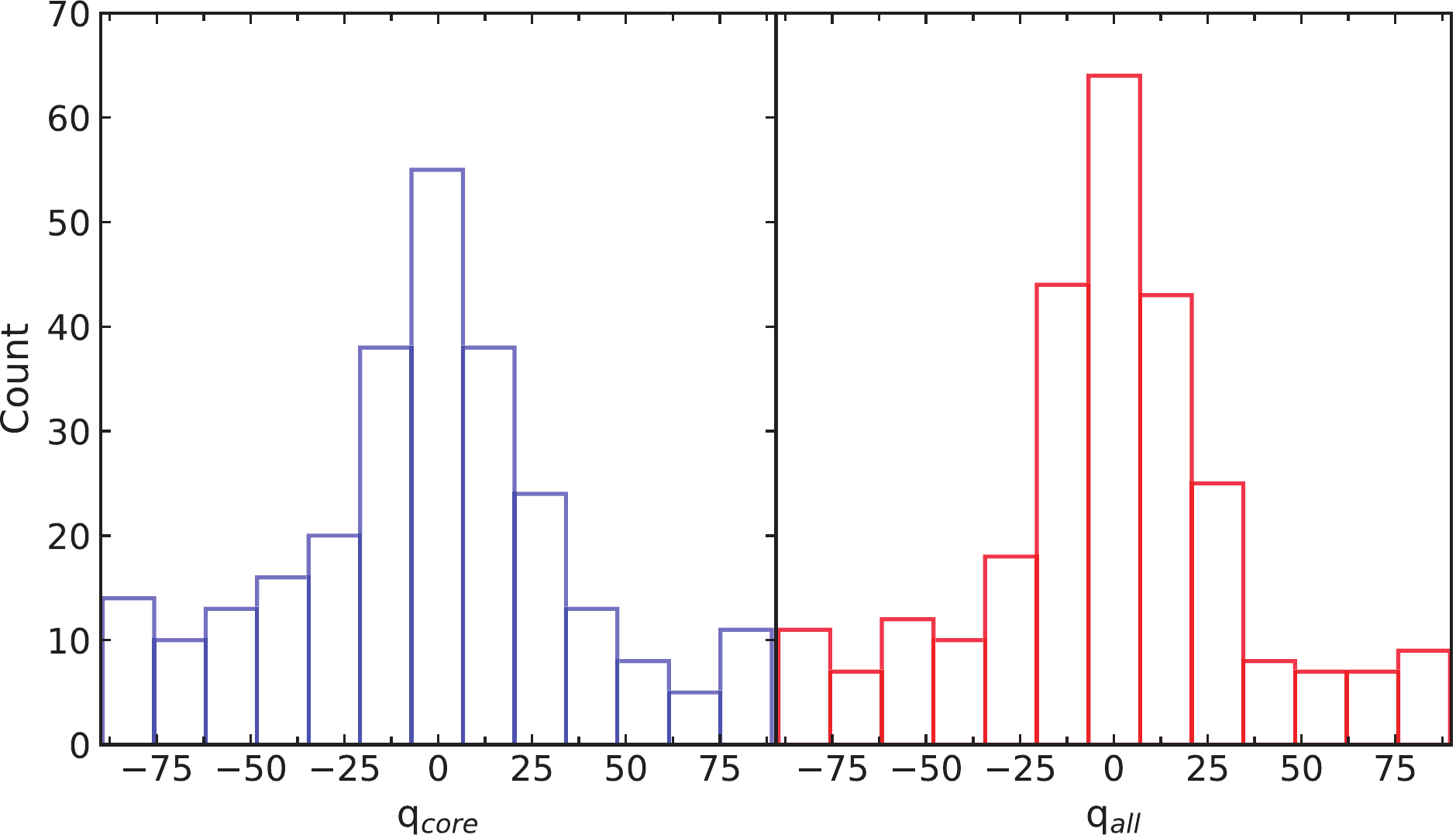}
\caption{Distribution of the orientations ($q_{core}$ and $q_{all}$) of this work. The blue histogram (left panel) corresponds to $q_{core}$ and the red histogram (right panel) to $q_{all}$ of the sample clusters.}
\label{fig:orientations}
\end{figure}

As shown in Figure~\ref{fig:Whole_sample}, the ellipticities and orientations of the cores and overalls of the whole sample clusters are presented in the Galactic Coordinate System. The clusters in our sample are distributed around the Galactic plane within the range of the Galactic latitude ($b$) from -40$^o$ to 40$^o$. The histograms of the orientations of the cluster's cores and overalls are presented in Figure~\ref{fig:orientations}, which shows that the orientations of most clusters in the sample are within $|q|$~$<$~45$^o$, which is basically consistent with the finding of \citet{zhai17}. This means that the sample of open clusters is more elongated in a direction parallel to the Galactic plane than perpendicular to it, consistent with the finding of \citet{berg01}.

\subsection{Correlations between Ellipticities and the Ages, Number of member stars, and Masses of the Clusters}

\begin{figure}[ht]
\centering
\includegraphics[width=76mm]{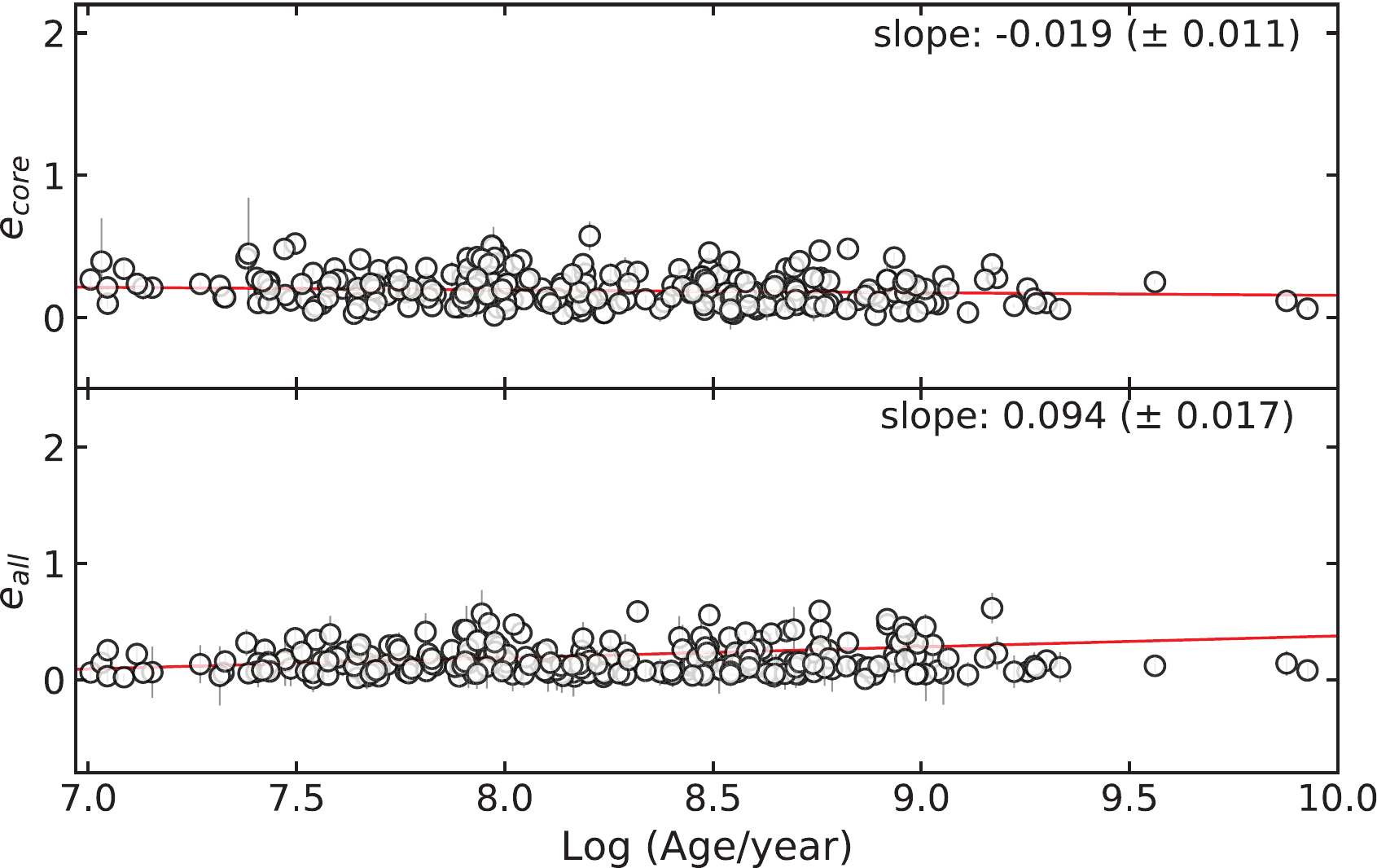}
\caption{The ellipticities ($e_{core}$ and $e_{all}$) as a function of the ages of the sample clusters. The inner ellipticities ($e_{core}$) are shown in the top panel, and the overall ($e_{all}$) in the bottom. The gray bars show the errors of the ellipticities. In each panel, the corresponding linear fit is shown as a red line.}
\label{fig:Age_ellipticities}
\end{figure}

With the ellipticities ($e_{core}$ and $e_{all}$) of the sample clusters obtained, we investigate the correlations between the ellipticities and the ages, the number of the member stars, and the masses of the clusters. Figure \ref{fig:Age_ellipticities} shows the ellipticities ($e_{core}$ and $e_{all}$) as a function of the ages of the sample clusters. The figure indicates that with age, the overall of the open clusters become more elliptical while their inner cores remains circular or slightly trend to circularize, which is consistent with what was inferred by \citet[][]{chen04, zhai17}. It should be expected that the core of a star cluster gets gradually circularised with age due to stellar dynamics, whereas the overall shape of a cluster continues to be elongated by external forces such as the Galactic tides, the Galactic differential rotation, or encountering with giant molecular clouds and so on. We should note that the significance of the linear fit to our sample is not strong, moreover, the shape parameters fitted possibly deviate from their true value due to the projection effect. More samples with three-dimensional information are needed to determine the true morphological evolution of the clusters over time.

\begin{figure}[ht]
\centering
\includegraphics[width=76mm]{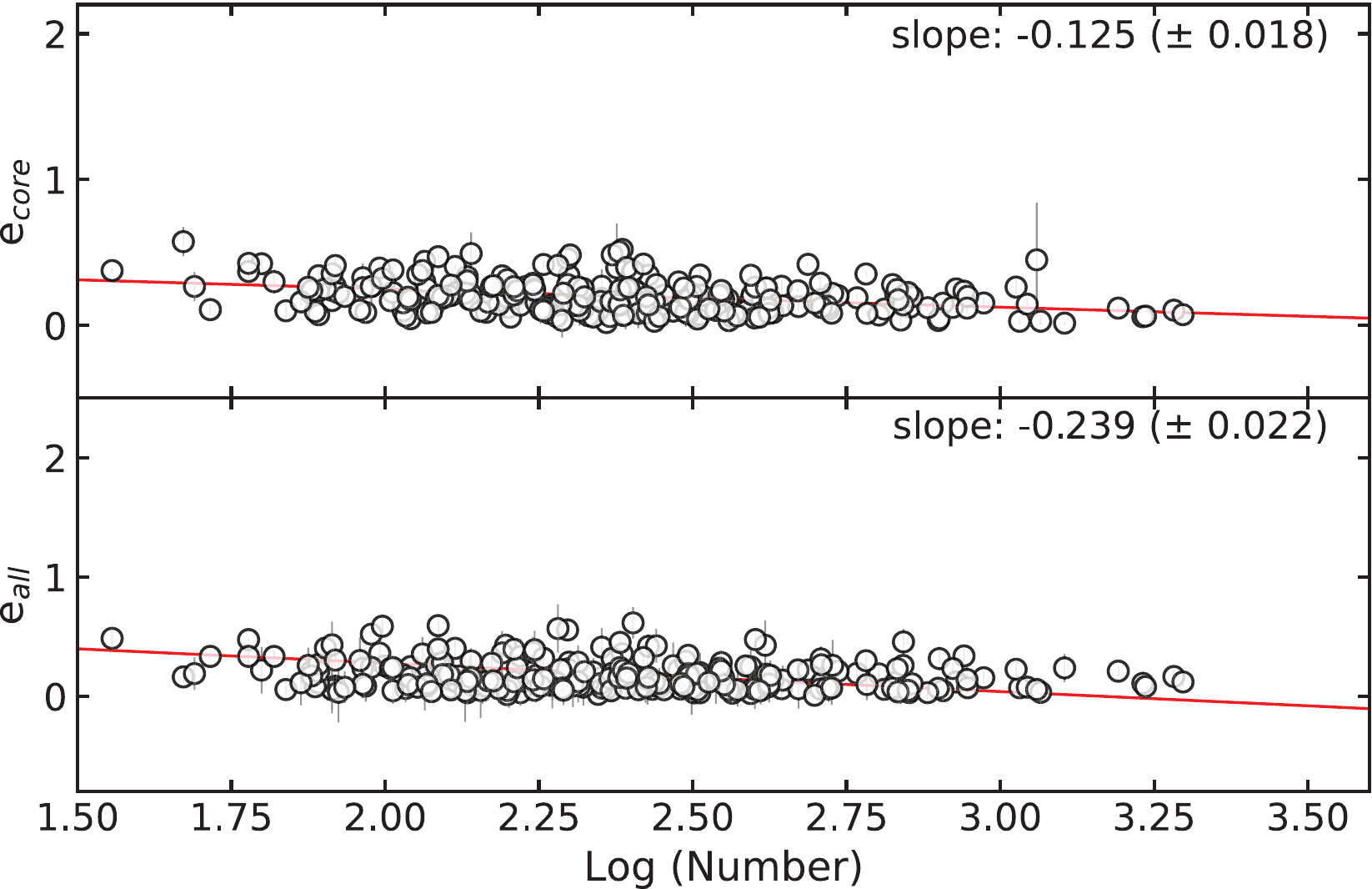}
\caption{The ellipticities ($e_{core}$ and $e_{all}$) as a function of the number of the member stars for the sample clusters. The symbols and bars in the two panels are the same as Figure \ref{fig:Age_ellipticities}. In each panel, the corresponding linear fit is shown by a red line.}
\label{fig:Member_ellipticities}
\end{figure}

Figure~\ref{fig:Member_ellipticities} shows the ellipticities of the sample clusters ($e_{core}$ and $e_{all}$) as a function of the number of member stars. From this figure, it can be concluded that both the inner ellipticities and the overall ellipticities of the clusters are negatively correlated with the number of the member stars of the clusters, which may imply that the higher the number of member stars of the clusters, the stronger the cluster's gravitational binding to resist the external disturbances, and the more it can keep smaller ellipticity. Therefore, clusters with more members should be able to go through a relatively stable shape evolutionary and thus survive for a long time in an arcane and complex dynamical environment. Conversely, the clusters with fewer members are more likely to disintegrate into field stars.

\begin{figure}[ht]
\centering
\includegraphics[angle=0,width=76mm]{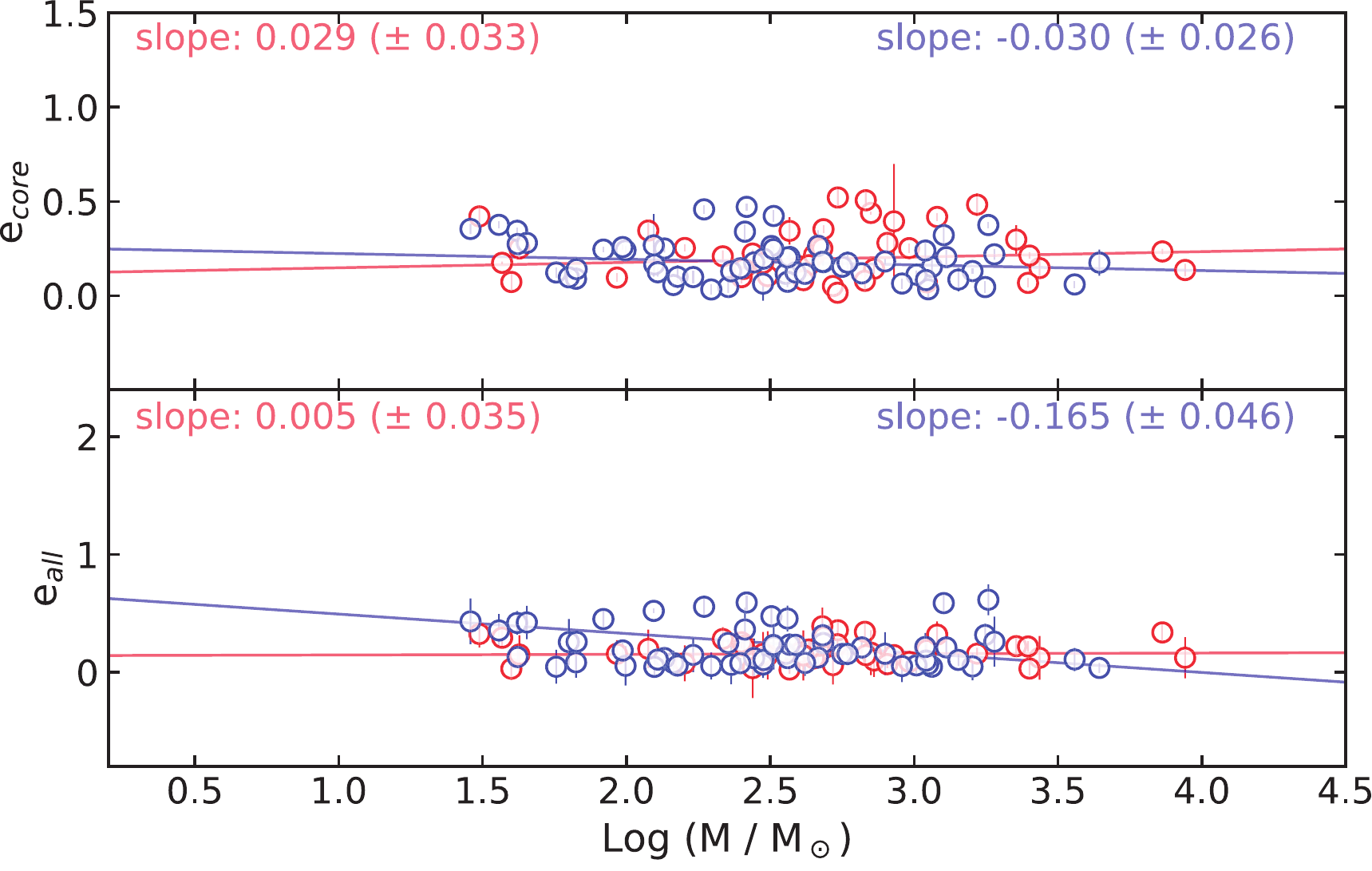}
\caption{The masses versus the ellipticities ($e_{core}$ and $e_{all}$) of 100 clusters in our sample. the subsample is divided into two age groups (log(age/year) $<$ 8 shown as red open circles and log(age/year) $\geq$ 8 as blue). In each panel, the corresponding linear fits are shown by the red and blue lines, respectively.}
\label{fig:Mass_ellipticities}
\end{figure}

\begin{figure*}[ht]
\centering
\includegraphics[angle=0,width=120mm]{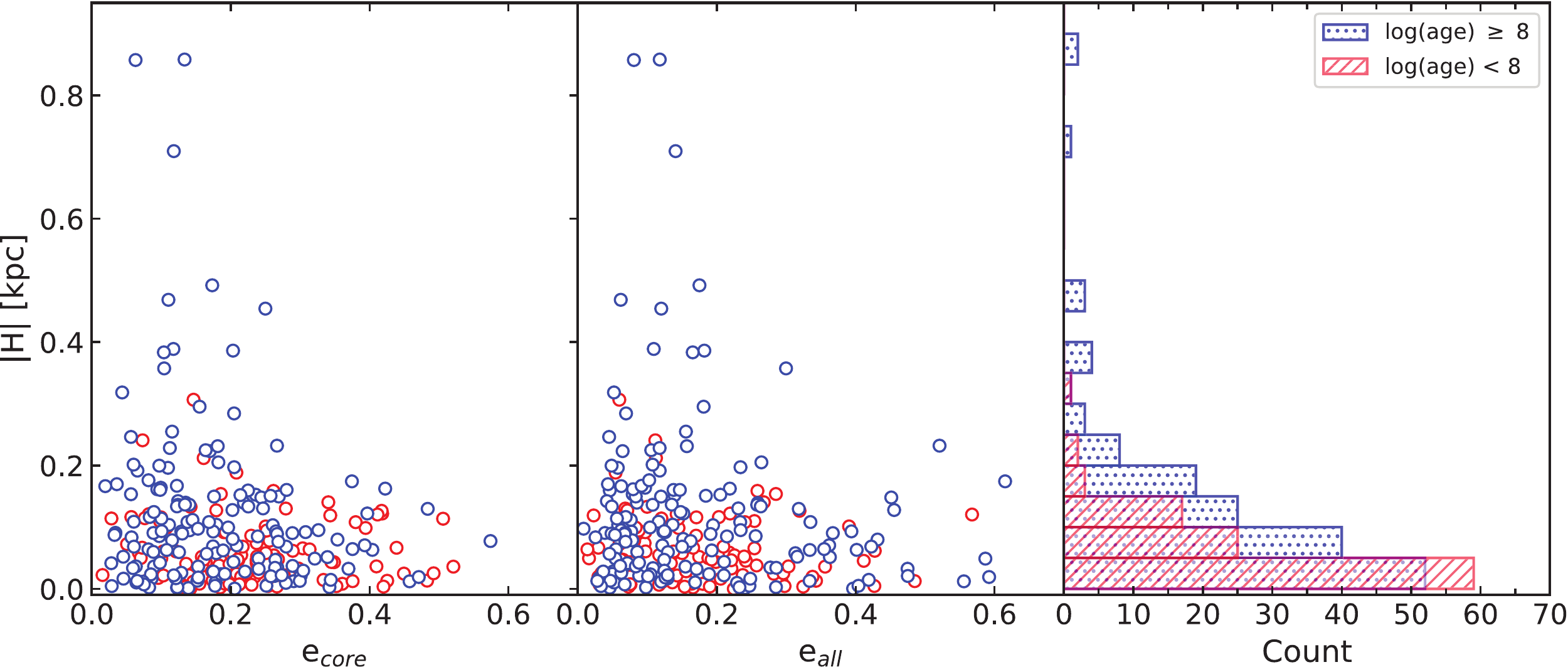}
\caption{The ellipticities ($e_{core}$ and $e_{all}$) vs. the heights from the Galactic disk for two age groups from the whole sample (the left and middle panels), and, the histogram of the heights of them (the right panel). The vertical axis of the three panels is the absolute value of the heights. The blue and red circles in the left and the middle panels correspond to clusters with log(age/year) $\geq$ 8 and log(age/year) $<$ 8, respectively. The red hatched histogram in the right panel corresponds to clusters with log(age/year) $<$ 8 and the blue shadows histogram for clusters with log(age/year) $\geq$ 8.}
\label{fig:Ellipticities_hight}
\end{figure*}

It is well known that cluster masses are an important factor affecting the shape of open clusters. \citet[e.g.][]{pisk07, pisk08} has given the masses of a number of clusters, which are determined based on the tidal radius. To investigate the correlation between ellipticities and cluster masses, we obtained the masses of 100 common clusters by cross-matching our sample with the one given by \citet[e.g.][]{pisk07, pisk08}, and then plot the distribution of the ellipticities versus masses for these 100 clusters, shown in Figure~\ref{fig:Mass_ellipticities}. Previous studies have shown that the majority of the survival star clusters are expected to be dissolved in some $10^8$ years \citep[e.g.][]{wiel71, pand86, berg01, bona06, yang13}. Therefore, we set to log(age/year) = 8 as a grouping age, then divided the 100 clusters into two age groups, and shows the distribution between the ellipticity and mass of the two groups in Figure~\ref{fig:Mass_ellipticities}. As shown by the figure, no significant correlations between the inner ellipticities and masses and between the overall ellipticities and masses are detected for the clusters with log(age/year)~$<$~8. However, a negative correlation between the inner and overall ellipticities with masses is detected for the clusters with log(age/year)~$\geq$~8, especially more significant between the overall ellipticities and the masses. This may imply that older clusters with larger masses may have a large enough self-gravitational potential to keep their overall ellipticity small and thus maintain a stable shape evolution, especially when no giant molecular clouds or other events are encountered.

To investigate why there is no significant correlation between cluster ellipticity and its mass for log(age/year)$<$ 8 in Figure~\ref{fig:Ellipticities_hight}, we divided the whole sample of 265 clusters into two age groups of log(age/year) $\geq$ 8 and log(age/year) $<$ 8, and then plotted the ellipticities versus the absolute values of the heights from the Galactic disk. The left and middle panels of Figure~\ref{fig:Ellipticities_hight} show the ellipticities ($e_{core}$ and $e_{all}$) versus the heights for the two age groups, respectively, and the right panel of the figure show the histogram of the heights of them. As seen in the right panel of Figure~\ref{fig:Ellipticities_hight}, the heights of the older group are clearly more widely distributed than that of the younger group, with a considerable number of the clusters in the older group distributed at relatively high heights. This suggests that the younger group clusters are closer to the disk and maybe more disturbed by the gravity of the Galactic disk and the surrounding environment, therefore, the mass of the younger group clusters may not be the dominant factor affecting the cluster ellipticity, so the correlation between the cluster ellipticity and mass is not obvious, as shown in Figure 6. However, for the older clusters, they are more widely distributed in the height from the disk, and a considerable number of clusters are located in the higher height from the disk compared to the younger group, thus these clusters may be relatively less disturbed by the gravity of Galactic disk and the surrounding environment compared with the young group, so they show a weak negative correlation between the cluster ellipticity and mass, as shown in Figure~\ref{fig:Mass_ellipticities}. There is no significant difference in the inner ellipticities distribution between the two age groups (see the left panel of Figure~\ref{fig:Ellipticities_hight}). However, the overall ellipticities distribution of the samples with log(age/year) $\geq$ 8 is significantly wider than that of the younger group (see the middle panel of Figure~\ref{fig:Ellipticities_hight}), which indicates that the overall ellipticities maybe tend to widen with time evolution. The above distributions are consistent with the results of the distribution of the inner ellipticities and the overall ellipticities with cluster's ages shown in the top and bottom panels in Figure~\ref{fig:Age_ellipticities}, respectively.

\begin{sidewaystable}[ht]
\centering
\caption{The Parameters of 265 Open Clusters in This Paper}
\label{table:data1}
\small
\begin{tabular}{cccccccccccccccc}
\hline\noalign{\smallskip}
\hline\noalign{\smallskip}
(1) & (2) & (3) & (4) & (5) & (6) & (7) & (8) & (9) & (10) & (11) & (12) & (13) & (14) & (15) & (16)\\
Name & $l$ & $b$ & log(age/year) & H & Number & $e_{core}$ & $\sigma_{core}$ & $q_{core}$ & $e_{all}$ & $\sigma_{all}$ & $q_{all}$ & $a_{core}$ & $b_{core}$ & $a_{all}$ & $b_{all}$\\
-- & (degree) & (degree) & -- & (pc) & -- & -- & -- & (degree) & -- & -- & (degree) & -- & -- & -- & --\\
\hline\noalign{\smallskip}
Alessi1& 123.255 & -13.335 & 8.935 & -162.5 & 63 & 0.423 & 0.034 & 17.388 & 0.219 & 0.196 & -10.205 & 0.310 & 0.179 & 0.691 & 0.540\\
Alessi2& 152.333 & 6.365 & 8.759 & 68.5 & 276 & 0.280 & 0.026 & -12.179 & 0.423 & 0.141 & 6.858  & 0.529 & 0.381 & 1.962 & 1.133\\
Alessi5& 288.058 & -1.966 & 7.723 & -13.6 & 308 & 0.174 & 0.007 & -29.674 & 0.293 & 0.087 & -1.436  & 0.435 & 0.359 & 1.327 & 0.938\\
Alessi6& 313.643 & -5.544& 8.777 & -85.5 & 294 & 0.097 & 0.010 & -6.524 & 0.256 & 0.196 & 1.431 & 0.280 & 0.252 & 1.078 & 0.802\\
Alessi8& 326.501 & 4.259 & 8.137 & 49.2 & 82 & 0.240 & 0.022 & -0.251 & 0.054 & 0.167  & 8.270 & 0.249 & 0.189 & 0.682 & 0.646\\
Alessi10& 31.597 & -21.058 & 7.746 & -158.8 & 75 & 0.262 & 0.013 & -6.236 & 0.259 & 0.149 & -10.651 & 0.250 & 0.185 & 0.848 & 0.629\\
... & ... & ... & ... & ... & ...  & ... & ... & ... & ... & ... & ... & ... & ... & ... & ...\\
\hline\noalign{\smallskip}
\end{tabular}
\flushleft
\tablecomments{The table includes the following information of each cluster in the sample: (1) the cluster name (Name), (2) the Galactic longitude in degree ($l$), (3) the Galactic latitude in degree ($b$), (4) the logarithm of the cluster age in years (log(age/year)), (5) the height from the Galactic disk in parsecs (H), (6) the number of the member stars of each cluster (Number), quoted from other literature. Columns (7) and (10) are the ellipticities ($e_{core}$ and $e_{all}$) of the sample clusters. Columns (8) and (11) are the errors of corresponding to the ellipticities, respectively, ($\sigma_{core}$ and $\sigma_{all}$). Columns (9) and (12) are the orientations of corresponding to the two parts in degree ($q_{core}$ and $q_{all}$). Columns (13) and (14) are half-length and half-short axes of the inner ellipses, respectively. Columns (15) and (16) are half-length and half-short axes of the overall ellipses, respectively. The complete table is available online.}
\end{sidewaystable}

\subsection{The Stratification degree of Hierarchical Structure}

\begin{figure}[ht]
\centering
\includegraphics[angle=0,width=70mm]{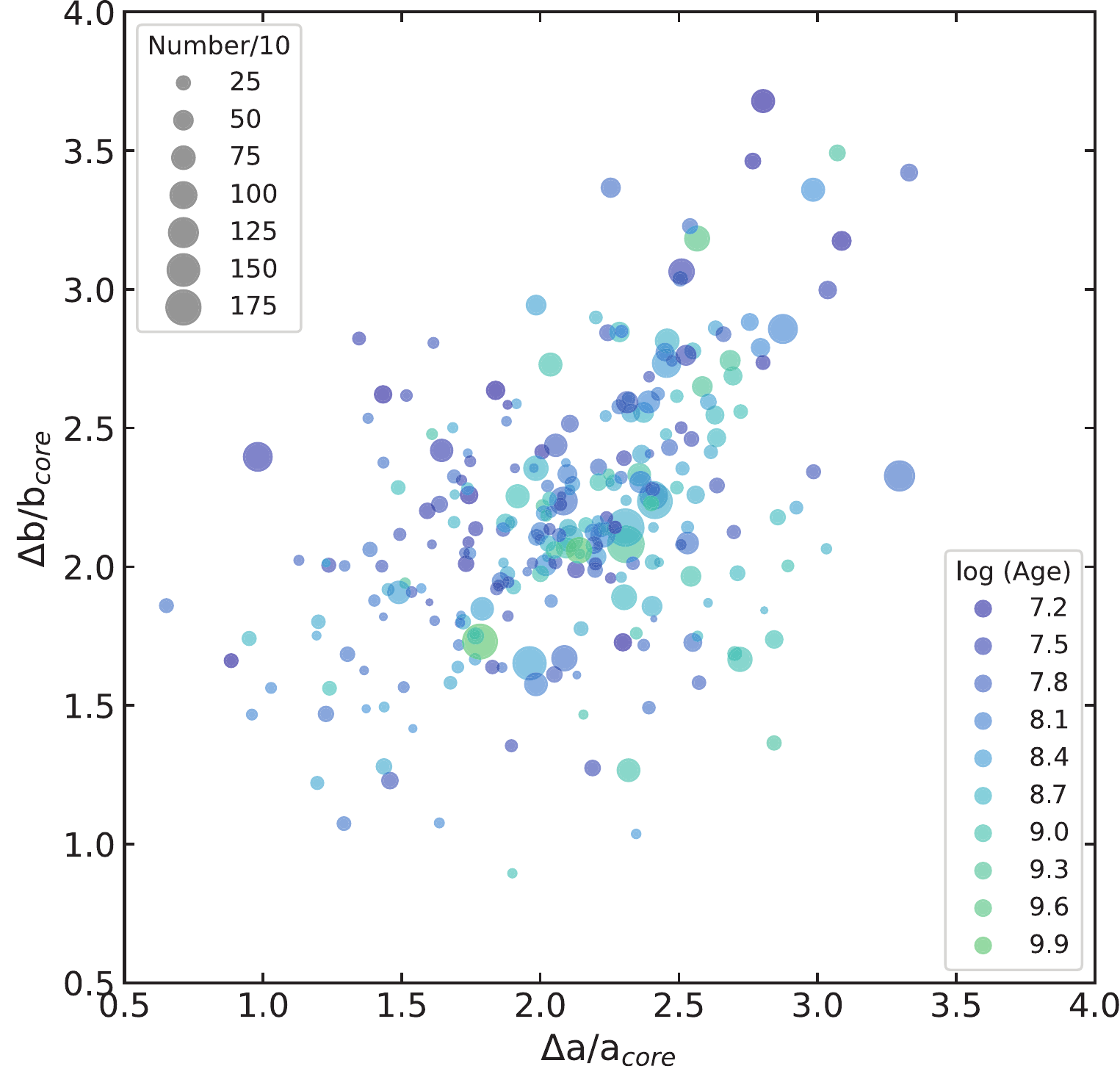}
\caption{The distribution of $\Delta a/a_{core}$ versus $\Delta b/b_{core}$ of the sample clusters. The color shade of the solid circles indicates the logarithmic ages (years) of the sample clusters. The size of the symbol is proportional to the number of sample cluster members.}
\label{fig:Radial_stratification}
\end{figure}

The hierarchical structure of open clusters has been verified \citep[e.g.][]{rabo98a, rabo98b, chen04}. But its detailed analysis, such as radial and tangential stratification degree of cluster interior, has been lacking. In our work, we use shape parameters to study the stratification degree. We define ($\Delta a/a_{core}$) and ($\Delta b/b_{core}$) as the radial stratification degree inside the cluster. In detail, ($\Delta a/a_{core}$) denotes the radial stratification degree in the long-axis direction, and ($\Delta b/b_{core}$) denotes the radial stratification degree in the short-axis direction. And $\Delta a$ and $\Delta b$ equal to $a_{all}-a_{core}$ and $b_{all}-b_{core}$, respectively. Moreover, we define $q_{all}-q_{core}$ as the tangential stratification within the cluster.

Figure~\ref{fig:Radial_stratification} displays the distribution of ($\Delta a/a_{core}$) versus ($\Delta b/b_{core}$), which illustrates the radial stratification degree within the cluster. It can be seen from this figure that for those clusters with a high number of member stars, the radial stratification degree in the short axis direction is almost the same as that in the long. The radial stratification range of most clusters with a small number of member stars is larger than that of clusters with a large number of member stars, both in the short-axis direction and in the long-axis direction within the clusters. In addition, we find that the age of the sample clusters is possibly related to the degree of radial stratification. To explore this issue, we define $p$ as the ratio of the radial stratification degree in the short axis direction to that in the long axis direction. In detail, $p$~$>$~1 shows the radial stratification degree in the short axis direction is greater than that of the long axis, while the opposite is true for $p$~$\leq$~1. These parameters have complied in Table~\ref{table:data2}. For the sample clusters, we find that in about 59\% of the clusters, the radial stratification degree of the short axis is greater than that of the long. Furthermore, if we count $p$ by different age groups, finding that in about 67\% of young clusters in (log(age/year)~$\leq$~8), the radial stratification degree in the short axis direction is greater than that in the long-axis direction, while for the remaining age groups, about half of the clusters in each age group have greater radial stratification degree in the short-axis direction than in the long-axis. This manifests that the number of clusters in the sample with a greater radial stratification in the short-axis direction than in the long-axis direction is almost the same as the opposite case. However, for most young sample clusters, the radial stratification degree of the short axis direction is greater than that of the long axis.

Strictly speaking, however, we should consider the shape parameters we derive as lower limits on the true shape parameters of the cluster. As long as a cluster has the form of a triaxial ellipsoid, we observe the projection of the cluster rather than its true size. The degree of possible discrepancy depends on the axial relationship of the ellipsoid and the direction of the axis relative to the cluster's line of sight \citep{pisk08}. It is worth noting that the true shape of clusters in space is currently unknown, and there may be projection effects in our results. Nonetheless, some plausible conclusions can be drawn, in particular, for young clusters located at low Galactic latitudes, the projection effect may have relatively little impact on these clusters. Therefore, the result of the radial stratification degree in the short axis direction being greater than that in the long-axis direction for most young clusters may provide clues to reveal how gas dissipation affects the morphology of young clusters' interiors. Since a young cluster with an age of only a few million years does not have enough time to completely change its internal structure through dynamical evolution, it is possible that the internal structure of the cluster may be affected by the gas dissipation processes, especially in the direction of the short axis of the cluster interior.

\begin{figure*}[ht]
\centering
\includegraphics[angle=0,width=140mm]{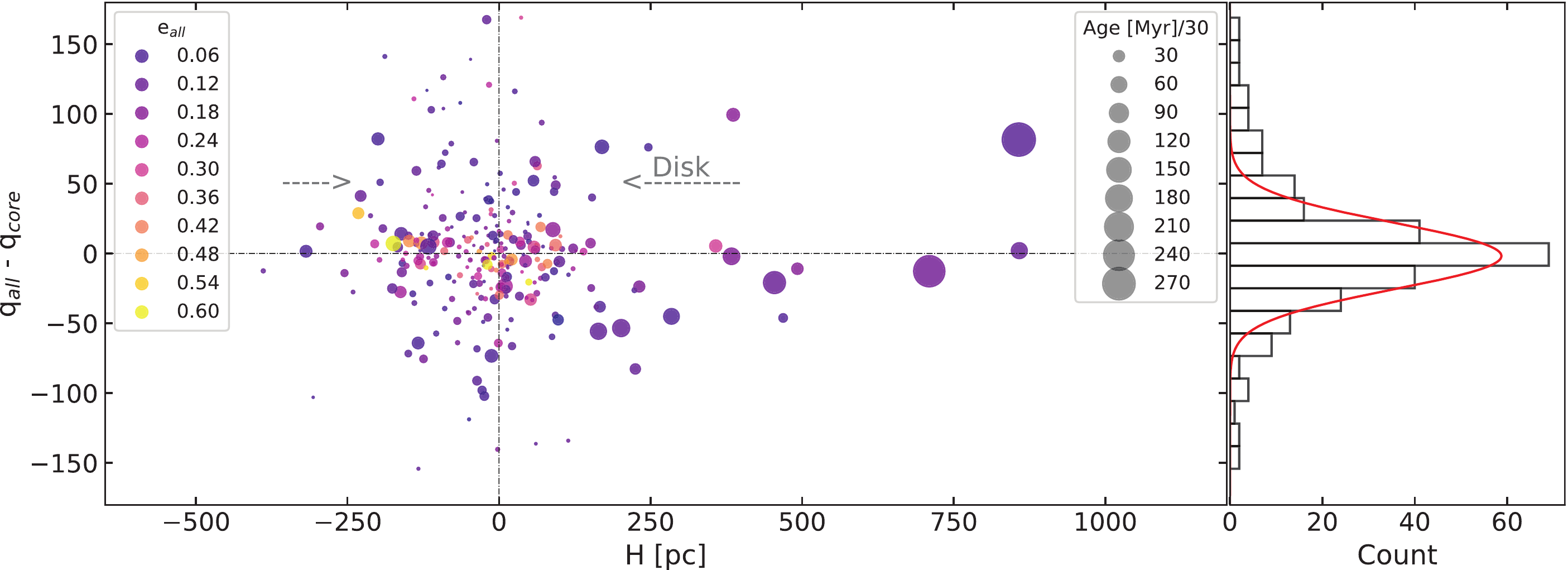}
\caption{The heights (H) from the Galactic disk versus $q_{all}$-$q_{core}$ of the sample clusters (left panel), as well as the histogram of the orientation difference ($q_{all}$-$q_{core}$) of them (right panel). The size of each circle in the left panel indicates the age of a cluster, with its color displaying the overall ellipticities ($e_{all}$). The vertical dash line in the left panel represents the Galactic plane, and the horizontal dash line in this panel indicates the zero stratification degree at the tangential direction of the cluster interior. The red curve in the right panel is the gauss fit to the histogram.}
\label{fig:Tangential_stratification}
\end{figure*}

We also investigate the tangential stratification in the interior of the cluster. From the right panel of Figure~\ref{fig:Tangential_stratification}, we can see that the tangential stratification inside the clusters shows a concentrated distribution. The left panel of Figure~\ref{fig:Tangential_stratification} shows that the open clusters with large overall ellipticity have relatively low tangential stratification, while clusters with small overall ellipticity near the Galactic plane are mostly in the opposite state. Older open clusters may exhibit lower stratification in the tangential direction. However, the number of old clusters is relatively small in the sample. This trend needs to be further verified by expanding the sample of old clusters.

The left panel of Figure~\ref{fig:Tangential_stratification} also shows that the young clusters are mainly distributed near the disk, and their tangential stratification covers almost the entire vertical axis. Open clusters still embedded in their natal clouds are called embedded clusters \citep{pelu12}. They evolve into optically observed open clusters at the cost of losing mass \citep[called ``infant losing weight'';][]{good06}.  It is important to note that these young clusters are vulnerable to self-evolution and external forces. As a result, they tend to exhibit a high degree of stratification. Furthermore, when approaching the Galactic disk, the external environment becomes more arcane and complex, and naturally, the degree of stratification should likely be high. However, the young clusters in the sample are almost uniform in the distribution of the tangential stratification degree. This needs to be further investigated in depth. Furthermore, it is intriguing why clusters with large overall ellipticity have little tangential stratification. Perhaps, this may be related to the dynamic evolution and the struggle of external forces, the exact nature of which is also worthy of further investigation.

It is expected that in the future, as the data from the major survey telescopes become more accurate and homogeneous, we can obtain the real distribution of star clusters in three-dimensions with the help of high-precision single-star distances, thus allowing a deeper understanding of the stratification of open clusters.

\begin{table}[ht]
\caption{Grouping Data}
\centering

\label{table:data2}
\begin{tabular}{c c c c c}
\hline\noalign{\smallskip}
\hline\noalign{\smallskip}
\hspace{0cm}&logA $>$ 0 & logA $\leq$ 8 & 8 $<$ logA $\leq$ 9 & logA $>$ 9 \\
\hline\noalign{\smallskip}
$p$ $>$ 1 &157 & 72 & 74 & 11 \\

$p$ $\leq$ 1 &108 & 35 & 65 & 8 \\
\hline\noalign{\smallskip}
\end{tabular}
\flushleft
\tablecomments{$p$ is defined as the ratio of the radial stratification degree of the short-axis to the radial stratification degree of the long-axis. The logA represents the log(age/year). These numbers in the table are the number of clusters from four different age groups.}
\end{table}

\subsection{Influence of Galactic Tides and Differential Rotation on Cluster Morphology}

We use the overall ellipticities of the clusters at two different directions in the Galactic plane to trace the influence of the Galactic tides and differential rotation on the cluster morphology. The following coordinate systems for the location of open clusters are applied: the spherical Galactic coordinates $(l, b)$; the XYZ coordinates in a Galactic reference frame centered on the Sun where these position parameters $(X, Y, Z)$ are directly obtained from \citet{cant18}. Since the cluster morphology is perpendicular to our line of sight, when the sight direction is at $l = 90^o$ or $270^o$, we see their (X, Z) projection (Z-axis from the Galactic center to North Galactic pole), revealing the full elongation toward the Galactic center. It is assumed that the elongation of the cluster morphology is mainly caused by the Galactic tides toward the Galactic center. However, at $l = 0^o$ or $180^o$, we see their (Y, Z) projection, revealing the elongated in the direction of Galactic differential rotation. The cluster morphology elongated may be mainly caused by the shear forces that existed at the direction of the Galactic differential rotation. Therefore, the projection effect is used to study the influence that the Galactic differential rotation and the Galactic tides toward the Galactic center exert on the morphology of the clusters.

\begin{figure}[ht]
\centering
\includegraphics[angle=0,width=90mm]{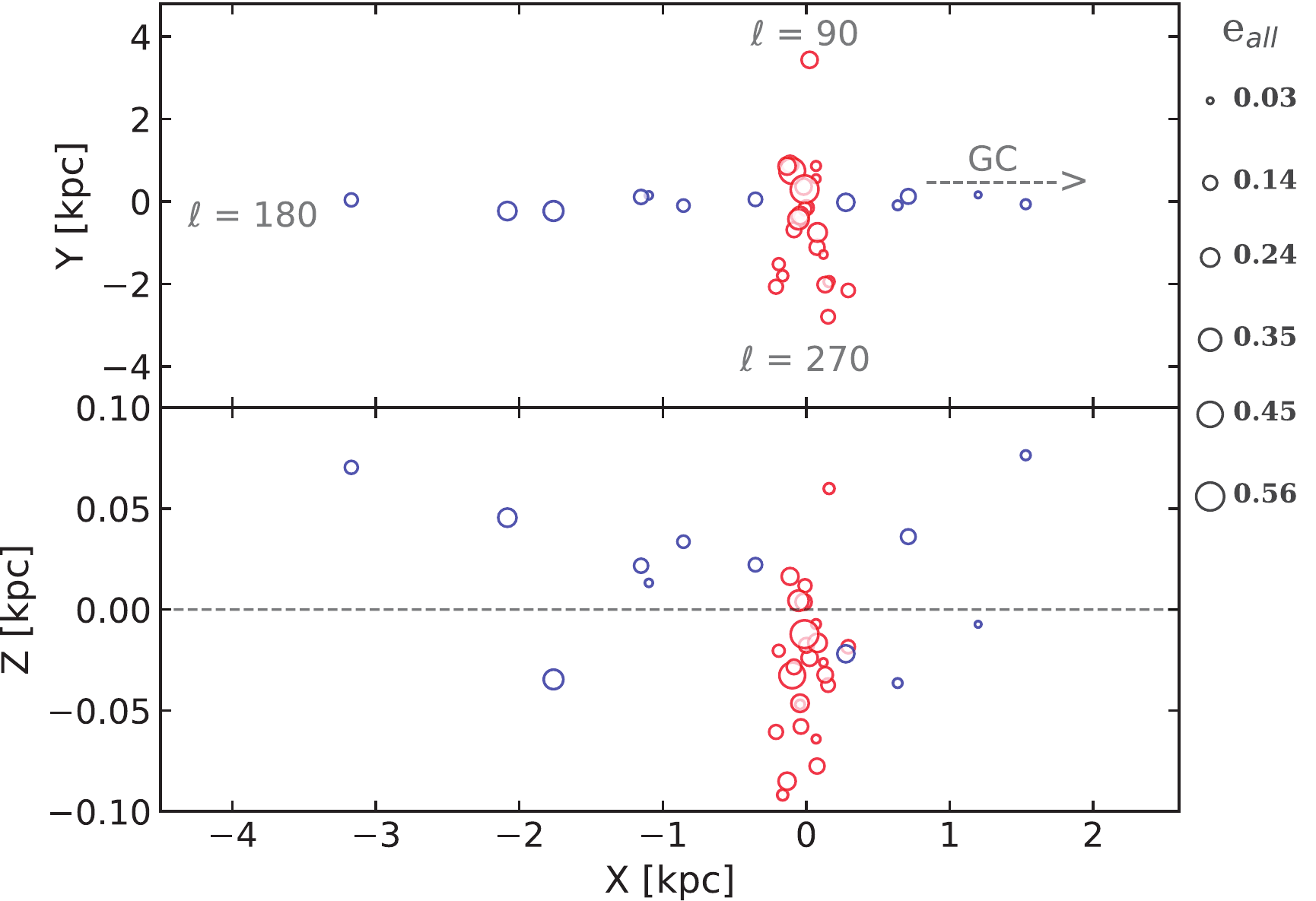}
\caption{The distribution of two subsamples elected from the whole sample in Galactic disk (X-Y plane, top panel) and X-Z plane (bottom panel). The red open circles in the top and bottom panels indicate the clusters at the tangential direction ($(80^o \leq l \leq 100^o) $ and $(260^o \leq l \leq 280^o)$) within $|Z|$ $\leq$ 100~pc, and the blue at the radial ($(170^o \leq l \leq 190^o)$ and $(350^o \leq l \leq 10^o)$) within $|Z|$ $\leq$ 100~pc. The sizes of the open circles in the two panels are proportional to the overall ellipticities of the subsamples. The black dash line in the bottom panel represents the Galactic disk. The scales of the circles are shown on the right side of the panel.}
\label{fig:Distribution_subsample}
\end{figure}

We divide the region around the sun into two parts: radial ($(170^o \leq l \leq 190^o)$ and $(350^o \leq l \leq 10^o)$) and tangential ($(80^o \leq l \leq 100^o) $ and $(260^o \leq l \leq 280^o)$) parts. And, the clusters in both parts within $|Z|$~$<$~100~pc are selected as two subsamples, i.e. the radial and tangential subsamples. And then the two subsamples are both displayed in the Galactic plane ($X-Y$ plane) and $X-Z$ plane in Figure \ref{fig:Distribution_subsample}. It can be seen from the figure that the overall ellipticities of most of the clusters in the tangential subsample are larger than that of the clusters in the radial. We calculate the average of the overall ellipticities of the 265 clusters and that of the two subsamples separately. The average value of the overall ellipticities of the radial subsample is smaller than that of the whole sample, while that of the tangential subsample larger than that of the whole. Then, the ratio of the average differences of overall ellipticities between the radial subsample and the whole sample to that between the tangential subsample and the whole is calculated, and it is about -8. This means that the Galactic tides toward the Galactic center may have more influence than the shear forces embedded in the Galactic differential rotation on the overall ellipticities of our sample clusters.

\subsection{Evolutionary Distribution of the Sample Clusters}

\begin{figure*}[ht]
\centering
\includegraphics[angle=0,width=100mm]{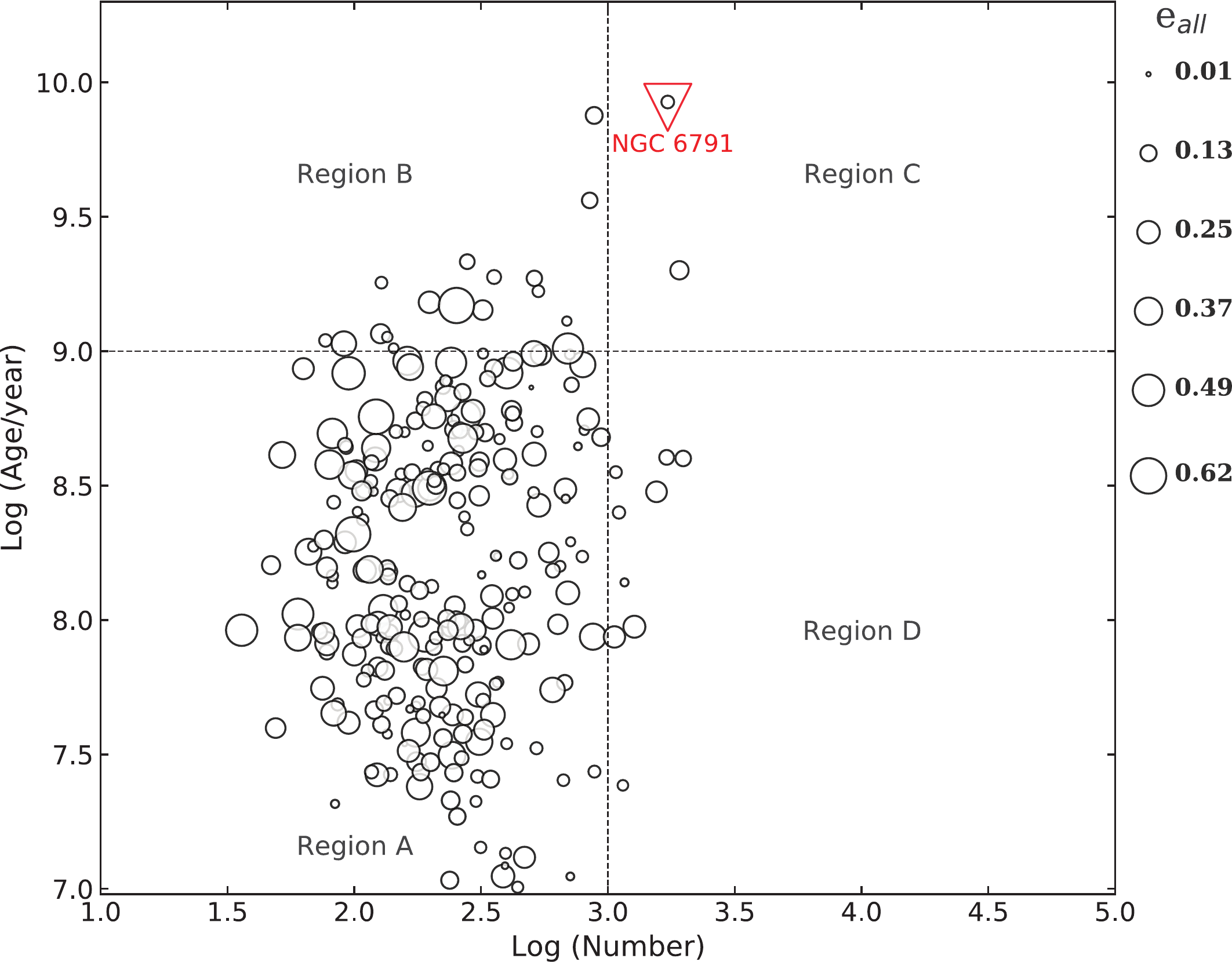}
\caption{The number of the member stars vs. the ages of the sample clusters. The sizes of the open circles in the panel are proportional to the overall ellipticities of the sample clusters. The red open triangle in the panel indicates the cluster (NGC~6791). Two black dash lines are the boundary lines of different partitions. The two partitions in the left half of the panel are marked as Region~B and Region~A from top to bottom, and in the right half are marked as Region~C and Region~D respectively. A legend to display the scales of those open circles is placed on the right side of the panel. Given that the definition of supercluster masses (massive ($>$ $10^5$ $M_{\odot}$) \citep{emig20}) and assuming that the average mass of every single star is about 1~$M_{\odot}$, the right endpoint of the x-axis in the panel is set to 5.}
\label{fig:Member_age}
\end{figure*}

Figure~\ref{fig:Member_age} shows the distribution of the ages and the number of member stars of the whole sample clusters. The sizes of the open circles in the panel are proportional to the overall ellipticities of the sample clusters. It is now generally accepted that the cluster will disintegrate in a few hundred million years, so we have drawn a horizontal dashed line at the possible upper limit of the disintegration time (1~Gyr) in this figure. A vertical dashed line is drawn at the position where the number of cluster members is 1000. The whole panel in Figure~\ref{fig:Member_age} is then divided into four regions, namely Regions A, B, C, and D. As seen in Figure~\ref{fig:Member_age}, most of our sample clusters are located in Region A, which confirms that open clusters are a class of stellar systems with a certain age range and a range of the number of member stars.

Open clusters lose their member stars as they evolve. Therefore, assuming different loss rates of member stars during the cluster evolution, the clusters seem to exhibit a trend of gradually increasing overall ellipticities from the lower right of Region A to the upper left of Region A, as shown in Figure~\ref{fig:Member_age}. This may suggest that the gravitational binding of the cluster decreases during cluster evolution with the losing of the cluster members, and the cluster morphology is susceptible to external disturbances, thus showing a tendency for the overall ellipticities to become larger. We should note that the loss rate of cluster member stars does not necessarily remain constant throughout the evolution process. As the number of cluster members decreases and the surrounding environment changes during evolution, the loss rate may also change. The loss rate of member stars in cluster evolution needs to be studied in depth in the future.

The relatively low number of clusters in region~B is consistent with the finding of a lack of older clusters with ages $t$~$>$~1~$Gyr$ by \citet{pisk18}. It is interesting to note the lack of clusters in the lower-left corner of Region~A, which could be the result of infant mortality \citep[e.g.][]{lada03, bast05, fall05, good06}. Previous studies have shown that most stars born in clusters do not remain there past the age of 10 million years, and only about 10\% of stars are observed in gravitationally bound clusters at this age \citep{lada03}. This may be one possible reason for the lack of clusters in the lower-left corner of Region~A.

There are two old and rich clusters in Region~C. One of them, NGC~6791, is the most populous and oldest cluster in our sample with a relatively small overall ellipticity. \citet{dale15} estimated the initial cluster mass of NGC~6791 to be $(1.5-4)$~$\times$~$10^5$ $M_{\odot}$ based on analytical studies and N-body simulations. It is very unlikely that such an old cluster will maintain a very low member loss rate during its long-term evolution, suggesting the initial cluster of it should locate in the right lower part of Region~D. Therefore, the superclusters, massive ($>$ $10^5$ $M_{\odot}$) and compact clusters \citep{emig20}, may give a possible evolutionary origin for NGC~6791.

\section{Summary}

In this study, we use the member catalog of open clusters available in the literature \citep{cant18} to give the shape parameters of 265 open clusters. Meanwhile, we conclude in the following for the analysis of cluster morphology evolution by the shape parameters:

1. The overall shapes of the sample clusters become more elliptical as they grow older, while their core shapes remains circular or slightly trend to circularize. The orientations of the most sample clusters are within $|q|$~$<$~45$^o$, which means that the sample clusters are more elongated in a direction parallel to the Galactic plane than perpendicular to it.

2. There is a negative correlation between the ellipticities and the number of member stars of the sample clusters, implying that the higher the number of member stars of the clusters, the stronger the cluster's gravitational binding to resist the external disturbances, and the more it can keep smaller ellipticity.

3. No significant correlations between the inner ellipticities and masses and between the overall ellipticities and masses are detected for the clusters with log(age/year)~$<$~8. However, negative correlations of the inner and overall ellipticities with masses are detected for the clusters with log(age/year)~$\geq$~8, especially more significant between the overall ellipticities and the masses, suggesting that the overall shapes of the clusters may be influenced by the cluster's mass, in addition to other factors.

4. For sample clusters with a high number of member stars, the radial stratification degree in the short axis direction is almost the same as that in the long. Both within the clusters in the short-axis direction and long-axis direction, most sample clusters with fewer members have a larger radial stratification range than those with more members. The number of clusters in the sample with a greater radial stratification in the short-axis direction than in the long-axis direction is almost the same as the opposite case. For most young sample clusters, the radial stratification degree of the short axis direction is greater than that of the long, implying that the degree of stratification in both directions within the young sample cluster might be unevenly affected by gas dissipation processes. The tangential stratification inside the clusters of our sample shows a concentrated distribution. The sample clusters with relatively large overall ellipticity have relatively low tangential stratification, while the clusters with smaller overall ellipticity near the Galactic plane are mostly in the opposite state. Older sample clusters may exhibit lower stratification in the tangential direction, implying that those clusters may continue to survive for a long time at a low level of stratification.

5. Our analysis of two subsamples of clusters distribution on the radial and tangential directions shows that the overall ellipticities of sample clusters may be more susceptible to the influence of Galactic tides toward the Galactic center than the shear forces embedded in Galactic differential rotation.

6. Sample clusters appear to show a general trend of increasing overall ellipticity from the lower right of Region A to the upper left of Region A, which means the loss of cluster members drives the overall cluster morphology to become elliptical. By analyzing the distribution of the ages and number of members of star clusters, we suggest that NGC~6791 may originate from superclusters.

We acknowledge that our sample clusters are still incomplete. A large and complete sample is needed to obtain more reliable conclusions about the shape evolution of open clusters. Besides, It is expected that in the future, as the data from the major survey telescopes become more accurate and homogeneous, we can obtain the real distribution of star clusters in three-dimensions, thus allowing a deeper understanding of the morphology evolution of open clusters.

\acknowledgments This work was supported by the National Natural Science Foundation of China under grants U2031209 and U2031204, the Youth Innovation Promotion Association CAS (grant No.2018080), 2017 Heaven Lake Hundred-Talent Program of Xinjiang Uygur Autonomous Region of China. The Authors thank to the reviewer for the very helpful comments and suggestions. We are grateful to Prof. Dr. Wen-Ping Chen for an in-depth discussion and suggestions on the collaboration. We would also like to thank Ms. Chunli Feng for touching up the language of the article. This study has made use of the Gaia DR2, operated at the European Space Agency (ESA) space mission (Gaia). The Gaia archive website is \url{https://archives.esac.esa.int/gaia/}.

\end{document}